# Predicting activation energies for vacancy-mediated diffusion in alloys using a transition-state cluster expansion

Chenyang Li,[1] Thomas Nilson,[1] Liang Cao,[1] and Tim Mueller[1, *]

[1]Department of Materials Science and Engineering, Johns Hopkins University, Baltimore, MD 21218, USA

[*] Email: tmueller@jhu.edu

**ABSTRACT**

Kinetic Monte Carlo models parameterized by first principles calculations are widely used to simulate atomic diffusion. However, accurately predicting the activation energies for diffusion in substitutional alloys remains challenging due to the wide variety of local environments that may exist around the diffusing atom. We address this challenge using a cluster expansion model that explicitly includes a sublattice of sites representing transition states and assess its accuracy in comparison with other models, such as the broken bond model and a model related to Marcus theory, by modeling vacancy-mediated diffusion in Pt-Ni nanoparticles. We find that the prediction error of the cluster expansion is similar to that of other models for small training sets, but with larger training sets the cluster expansion has a significantly lower prediction error than the other models with comparable execution speed. Of the simpler models, the model related to Marcus theory yields predictions of nanoparticle evolution that are most similar to those of the cluster expansion, and a weighted average of the two approaches has the lowest prediction error for activation energies across all training set sizes.

# I. INTRODUCTION

Atomic diffusion plays an important role in determining various material properties such as the structure and activity of nanocatalysts [1-5], the conductivity and stability of batteries [6-9] and fuel cells [10-13], and the electrical properties of semiconductor devices [14-17]. Atomic diffusion in materials often consists of hops between local minima on the potential energy surface, with atoms vibrating about these local minima in the time between hops. If the rates of these transitions are known, the simulation of diffusion can be significantly accelerated through the use of kinetic Monte Carlo (KMC) [18,19], which simulates only the discrete jumps between local minima rather than continuous time evolution of the system. The key to a KMC simulation is the rapid calculation of transition rates between local minima. By transition state theory [20], these rates depend exponentially on the activation energies of the hops, making the rapid and accurate calculation of activation energies critically important to the success of a KMC simulation.

Several methods for predicting activation energies in KMC simulations have been proposed and explored in the literature. These include the broken bond model [21-28] and a model that is similar to Marcus theory [29-31], which have been used in KMC simulations to study diffusion-related properties of materials. These methods are relatively simple and can be applied to a variety of material systems. However, the simplicity of these models limits their accuracy, and there is no way to systematically improve them if the predicted activation energies are not sufficiently accurate. An appealing alternative is to use systematically improvable machine-learned energy models such as cluster expansions [32-34] and interatomic potentials [35-42] for the rapid and accurate prediction of activation energies. As the cluster expansion is a discrete model explicitly designed to calculate the energies of local minima on the potential energy surface, it is particularly well suited for KMC.

The cluster expansion has become a valuable tool for studying atomic order and structure-property relationships in alloys [43-58], but it is rarely used to model diffusion [33,59,60]. Soisson and Fu [61,62], building off of work by Bouar and Soisson [63], developed a cluster-expansion-like lattice model to predict activation energies in which the energy of the transition state is a sum of nearest-neighbor pair interactions. In 2001, Van der Ven *et al.* demonstrated how cluster expansions could be used to predict activation energies in bulk $Li_xCoO_2$ by constructing two cluster expansions: a global cluster expansion of the local minima on the potential energy surface, and a local cluster expansion around the hopping atom that calculated a "kinetically resolved activation barrier" [33]. The combination of the two cluster expansions enabled the rapid calculation of activation energies. In a separate paper, Van der Ven *et al.* used a cluster expansion to predict the energies of both end states and local minima along the diffusion path [64]. To calculate the energies of transition states between local minima, they added a constant value to the higher-energy local minimum, where the constant value depended on whether or not the higher-energy site shared a face with a nearby occupied site. In 2010, Bhattacharya and Van der Ven adapted this approach for situations in which an atom at the intermediate site along the diffusion path is at a saddle point rather than a local minimum [65]. They accomplished this by using nudged elastic band method [66,67] to calculate the energies of structures in the training set in which the atom was at the intermediate site. This approach enabled them to calculate the energies of both the initial and transition states using a single global many-body cluster expansion.

Here, we present a method equivalent to this latter approach, in which transition states are explicitly included in a global cluster expansion as a set of sublattice sites, for modeling vacancy-mediated diffusion in Pt-Ni nanoparticles. The advantage to using a single global cluster expansion, rather than separate global and local cluster expansions, is that a single cluster expansion may be more compatible with general-purpose cluster expansion software packages and easier to extend to include correlated hops in which multiple atoms change sites at the same time. Pt-Ni nanoparticles have been widely studied as potential catalysts for the oxygen reduction reaction [1,68-71] and we present a challenging system for predicting activation energies due to the variety of local environments that exist both in the bulk and on the surface. On this system we find the cluster expansion approach yields significantly more accurate activation energies than alternative leading approaches with little to no increase in computational cost for the kinetic Monte Carlo simulation.

## II. METHODS

### A. DFT

Density functional theory (DFT) [72] calculations were performed using the Vienna Ab initio Simulation Package (VASP) [73] with the revised Perdew-Burke-Ernzerhof (RPBE) [74,75] exchange-correlation functional. The Pt_pv_GW and Ni PBE projector-augmented wave (PAW) [76] potentials were used. A single *k*-point at the center of the Brillouin zone was used for each nanoparticle. For bulk materials, the Brillouin zone was sampled using generalized Monkhorst-Pack grids generated by the *k*-point grid server [77] with a minimum distance of 46.5 Å between real space lattice points. Second-order Methfessel-Paxon smearing [78] with a width of 0.2 eV was used to set partial occupancies. The convergence criteria for the electronic self-consistency iteration and the ionic relaxation were set to be $10^{-4}$ eV and 0.03 eV/Å, respectively. The climbing image nudged elastic band (CI-NEB) [66,67] method was used to calculate the activation energies for atomic diffusion, and the calculations were considered to be sufficiently converged when the maximum force component perpendicular to the diffusion path was below 0.05 eV/Å.

## B. Cluster expansion

Cluster expansions are generalized Ising models that account for many-body interactions [32,79] and are commonly used to study atomic order in substitutional materials [47,54-56,80-82]. In this application, the arrangement of atoms in the material is expressed as a set of discrete sites that are occupied by different elements. In the cluster expansion constructed here, the sites are arranged on an fcc lattice and each site is occupied by a Pt atom, Ni atom, or vacancy. We fit the parameters of the cluster expansion, known as effective cluster interactions (ECI), to DFT-calculated energies using the Bayesian approach [83], which has been shown to effectively improve the accuracy of the cluster expansion for a given training set size [54,81,83]. Specifically, we assumed exponential decay in the width of the prior probability distribution with respect to the number of sites in each cluster and the maximum distance between sites.

To calculate activation energies for diffusion using the cluster expansion we added a sublattice of "transition state" sites halfway between nearest-neighbor sites (Figure 1a). Each of these transition state sites corresponds to a saddle point on the potential energy surface for a hop between neighboring sites. During a single diffusive hop, an atom starts at one of the fcc sites, moves to an adjacent transition state site, and then ends up in a nearest-neighbor fcc site (Figure 1b). By comparing the calculated energy of the system when the atom is at the transition state site with the energy of the system when the atom is in the initial fcc site, we directly predict the activation energy from the cluster expansion. To train the cluster expansion, the energies of the configurations where an atom is at a transition state site were obtained from DFT nudged elastic band calculations.

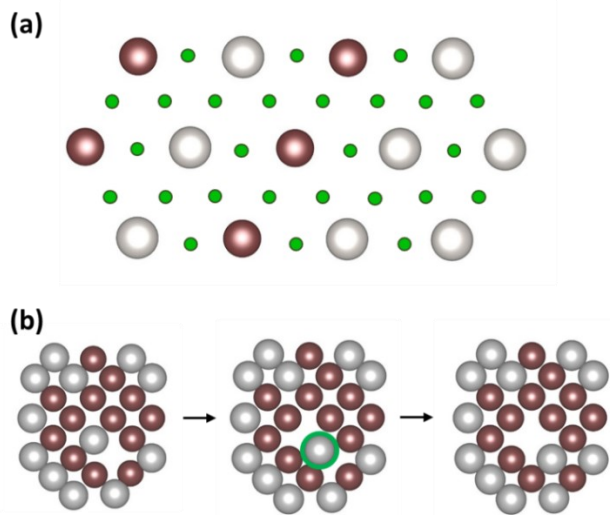

**Figure 1.** (a) A schematic of transition state cluster expansion. Large gray and brown spheres represent Pt and Ni, respectively, on lattice sites (local minima on the potential energy surface). Small green dots represent saddle points for hops between two local minima. For illustration purposes a two-dimensional lattice is shown, but the model in this paper is built on a three-dimensional fcc lattice. (b) A set of three training structures representing a diffusive hop, with the diffusing species at the transition state site circled in green. From left to right: initial, transition, and final state.

We constrained the set of allowed configurations to prevent simultaneous occupation of a transition state site and an adjacent fcc site, consistent with the mechanism of a vacancy-mediated hop. We also included a constraint that prevented two transition state sites from being occupied simultaneously, effectively allowing only a single atom to move during any hop. We note that this framework could be extended in future work by removing this second constraint. This would allow for collective diffusion, in which multiple atoms hop simultaneously, which is particularly likely to occur on surfaces [84-86].

The constraints on the set of allowed configurations make the cluster expansion overdetermined, i.e. there are more possible cluster functions than possible configurations. The cluster expansion can therefore be simplified by removing unnecessary functions. Here we accomplish this by removing all functions associated with clusters that include two sites that cannot be simultaneously occupied. As the number of remaining cluster functions is then equal to the number of allowed configurations, and these functions are linearly independent, they form a complete basis for the constrained configuration space. The proofs of these statements are provided in the Appendix. We note that this simplification procedure is not limited to studies of diffusion and can be used in any cluster expansion in which some sites cannot be simultaneously occupied.

A training set of 299 relaxed structures was generated using DFT for fitting the cluster expansion. The training set contained five bulk structures and the initial, transition, and final states of diffusive hops in different nanoparticles (Figure 1b). Each of the five bulk structures was included twice to ensure accuracy in the bulk limit. The cluster expansion was truncated to include the empty cluster, point clusters, all two-body clusters up to a cutoff distance of 8 Å, all three-body clusters up to a cutoff distance of 6 Å, all four-body clusters up to a cutoff distance of 4 Å, all five-body clusters up to a cutoff distance of 4 Å, and all six-body clusters up to a cutoff distance of 4 Å. The resulting cluster expansion contained 1097 symmetrically distinct cluster functions. In addition to the 299 energies in the training set, we also included 196 activation energies (energy differences between the transition state and the two corresponding end states) explicitly in the fitting to improve accuracy, as the expressions for these energies only include the terms that change between the initial and transition states. An additional regularization parameter was introduced to distinguish the expected magnitude of ECI for clusters that include and do not include transition states. Additional information about how the cluster expansion was fit can be found in section 1 of the Supplementary Material [87].

## III. RESULTS AND DISCUSSION

We begin by assessing the predictive accuracy of the transition-state cluster expansion on both formation energies and activation energies. The root-mean-square leave-one-out cross validation (LOOCV) errors are 1.203 meV/atom for the formation energies and 0.127 eV per particle for the activation energies (Figure 2), relative to DFT calculations. The LOOCV error of the activation energies is equivalent to 0.944 meV/atom, which is of the same order of magnitude as the error for the formation energies. This LOOCV error compares favorably to validation errors for other machine learning methods for predicting activation energies (Table S1 [35-42]), especially considering that the DFT data set contains hops in a wide variety of coordination environments, including both in the bulk and on the surface.

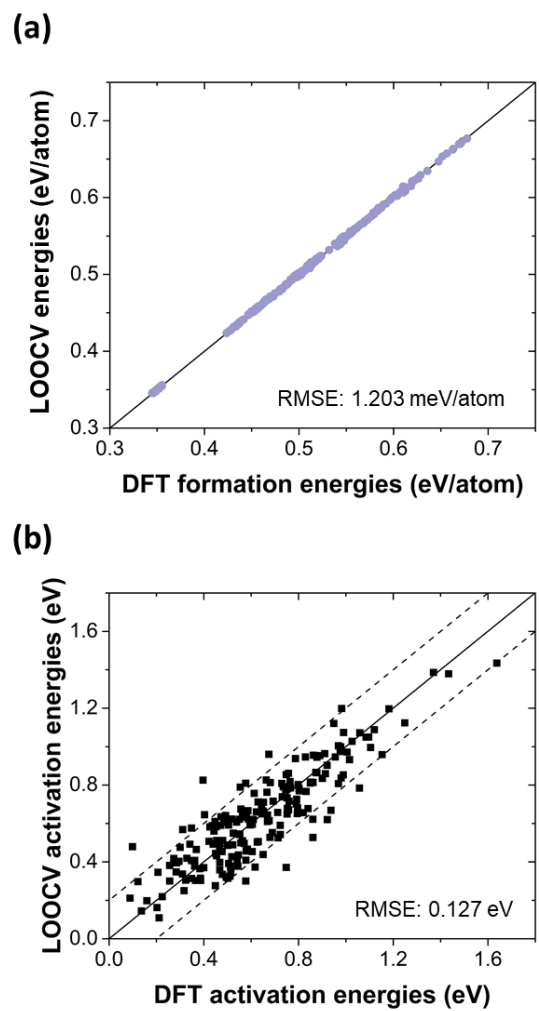

**Figure 2.** Leave-one-out cross validation (LOOCV) of a) the formation energies and b) activation energies from the transition-state cluster expansion. The dashed lines in (b) indicate $\pm\,0.2$ eV deviation from perfect agreement between DFT-calculated and cluster-expansion-predicted activation energies.

To further validate the accuracy of the transition-state cluster expansion for predicting the energies of local minima on the potential energy surface, we have used it to predict the equilibrium surface composition profile of a 4.5 nm cuboctahedral $Pt_3Ni$ nanoparticle ($Pt_{2547}Ni_{849}$) at 333 K (Figure S2, Supplementary Material section 2.1 [87]), using Metropolis Monte Carlo [88] simulations in a canonical ensemble. The near-surface composition profile (Figure S2) is similar to those which have been experimentally [68] and computationally [69,89] determined for an extended $Pt_3Ni(111)$ surface, consistent with the fact that the surface of the cuboctahedral Pt$_3$Ni nanoparticle consists of mainly (111) facets.

We have compared the accuracy of transition-state cluster expansion for predicting activation energies to three other simple models that have been developed for KMC simulations: the broken bond model [21], a model related to Marcus theory [29], and a model we have previously used in which the activation energy for a hop from a higher-energy state to a lower-energy state (i.e. a "downhill" hop) is a constant value [1,90]. Each of these models was trained using the same data used to train the cluster expansion. In each of these models other than the broken bond model, the cluster expansion is used to calculate the energy of each end state and the transition state energy is expressed as a function of the end state energies. The functional forms of the simple models we have evaluated are shown in Table 1.

In the broken bond model [21], the activation energy for diffusion is calculated as a linear function of the number of nearest-neighbor bonds of each type (e.g. Pt-Pt, Pt-Ni, or Ni-Ni) that are broken when the hopping atom leaves its initial state. One of the drawbacks of this model is that it does not satisfy detailed balance, as the difference in activation energies between the forward and reverse hops will generally not be the same as the difference in energies between the end states. We have also evaluated a broken-bond approach [24] in which detailed balance is restored by calculating a kinetically resolved activation (KRA) energy [33] as a function of the average number of bonds of each type at the initial and final states.

In the model related to Marcus theory [29], the potential energy surface of the initial and final state of the reaction as approximated using a simple parabolic form. In this model the activation energy can be analytically expressed as a function of the difference between the initial and final states,

ΔE (Table 1, Supplementary Material section 3 [87]). We will refer to this model as the "parabolic potential" model.

**Table 1.** Models used as comparisons to the cluster expansion in this work. The fitted parameters are determined by minimizing the RMS LOOCV error. $n^i$ and $n^f$ are the number of bonds at the initial and final states, respectively.

| | Functional forms of activation energy | Fitted parameters | Schematics |
|---|---|---|---|
| Broken bond | $E_a = n^i_{Pt-Pt}E_{Pt-Pt} + n^i_{Pt-Ni}E_{Pt-Ni} + n^i_{Ni-Ni}E_{Ni-Ni}$ | $E_{Pt-Pt} = 0.097$ eV <br> $E_{Pt-Ni} = 0.066$ eV <br> $E_{Ni-Ni} = 0.067$ eV | 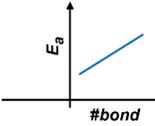 |
| Broken bond (KRA) | $E_a = \frac{1}{2}\Delta E + \frac{1}{2}(n^i_{Pt-Pt} + n^f_{Pt-Pt})E_{Pt-Pt}$ <br> $+ \frac{1}{2}(n^i_{Pt-Ni} + n^f_{Pt-Ni})E_{Pt-Ni}$ <br> $+ \frac{1}{2}(n^i_{Ni-Ni} + n^f_{Ni-Ni})E_{Ni-Ni}$ | $E_{Pt-Pt} = 0.097$ eV <br> $E_{Pt-Ni} = 0.067$ eV <br> $E_{Ni-Ni} = 0.066$ eV | |
| Parabolic potential | $E_a = \begin{cases} 0, & \Delta E < -4b \\ \frac{(\Delta E + 4b)^2}{16b}, & -4b \leq \Delta E \leq 4b \\ \Delta E, & \Delta E > 4b \end{cases}$ | $b = 0.628$ eV | 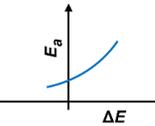 |
| Constant activation energy | $E_a = \begin{cases} c, & \Delta E < 0 \\ c + \Delta E, & \Delta E \geq 0 \end{cases}$ | $c = 0.510$ eV | 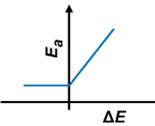 |
| Weighted average | $E_a = fE_a^{CE} + (1-f)E_a^{parabolic}$ | $f = 0.75$ | |

The final model, which we have used in previous work [1,90], uses a constant activation energy for hops to a lower-energy state and the same constant energy plus ΔE otherwise. The advantage of this model is that the value of the constant activation energy only affects the prefactor of the KMC simulation (Supplementary Material section 2.3 [87]), so it is possible to simulate the dynamics of diffusion without including any NEB-calculated activation energies in the training data. We will refer to this model as the "constant activation energy" model.

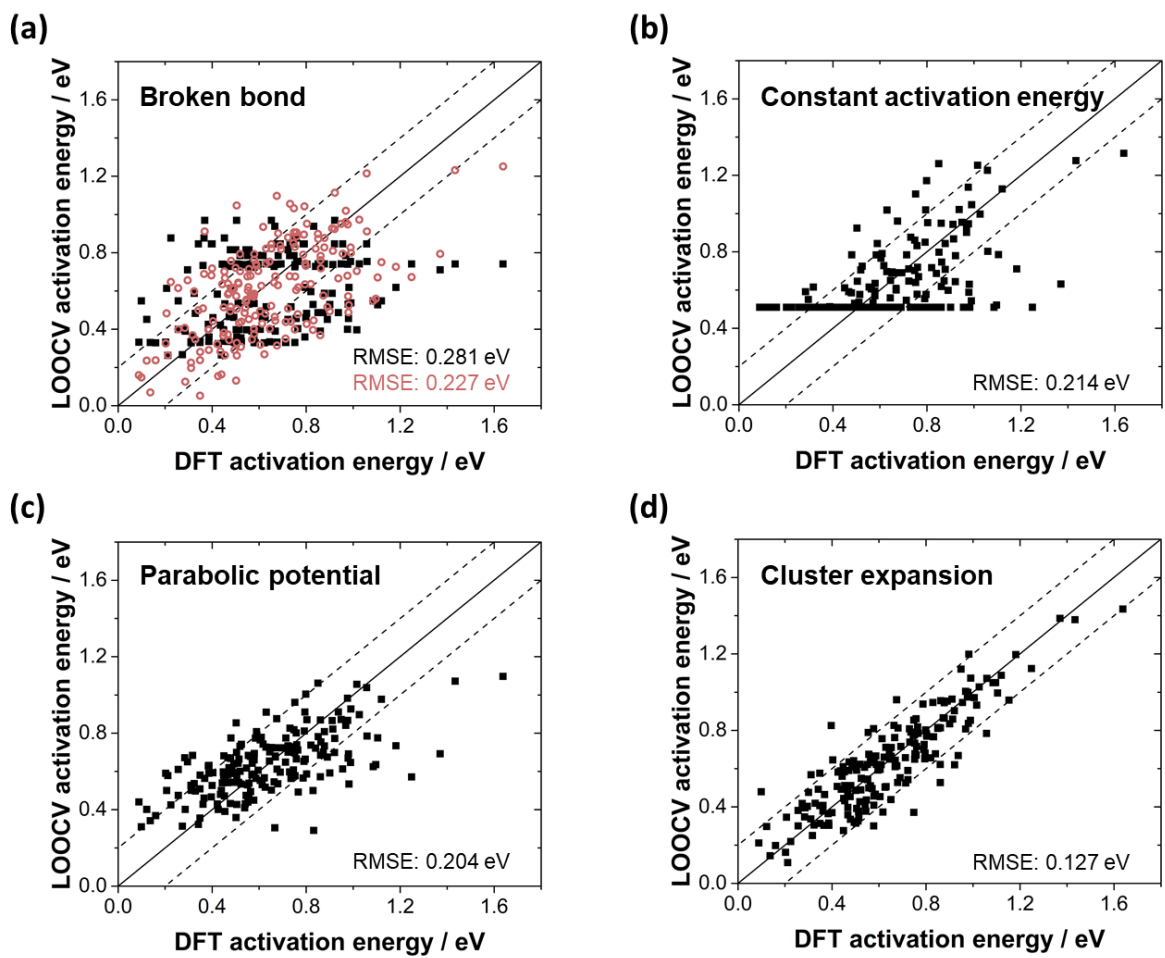

**Figure 3.** Leave-one-out cross validation (LOOCV) of the activation energy from the four methods against the known DFT activation energy in the training set. In a) the black and red data points are the simple broken bond and the KRA broken bond model, respectively. The dashed lines indicate $\pm 0.2$ eV deviation from perfect agreement.

The simple broken bond model has the largest root-mean-square LOOCV error (0.281 eV), followed by the KRA broken bond model (0.227 eV), the constant activation energy model (0.214 eV), the parabolic potential model (0.204 eV), and the cluster expansion (0.127 eV) (Figure 3). The relatively low error for the cluster expansion is due largely to the fact that it is systematically improvable. As a result, it includes far more parameters (1097) than the other models. Because the simple broken bond model only captures the nearest-neighbor environment around the initial state of the diffusing atom and its functional form establishes an upper bound for the activation energy (~1 eV), it predicts many hops with identical activation energies regardless of the end states and it underestimates the activation energies for many hops with large activation energies. The KRA broken bond model is significantly more accurate, likely because it does take into account the end states and makes the forward and reverse hops consistent. In particular, because this model depends explicitly on the energy difference between the initial and final states, it does not impose an upper bound on the activation energy as the broken bond model does. The constant activation energy model establishes a lower bound of the activation energy that is assigned to all downhill hops (Figure 3b). This causes overestimation and underestimation of small and large activation energies, respectively. The trend for the parabolic potential model (Figure 3c) is better, although it still significantly overestimates and underestimates some small and large activation energies respectively. A comparison of how the constant activation energy model and the parabolic potential models predict the activation energy as a function of reaction energy is provided in Figure S3.

It is often possible to create a linear combination of multiple models to create a new model with improved predictive accuracy, an approach generally known as ensemble learning [91]. To evaluate this approach, we created a model consisting of a weighted average of the cluster expansion and the parabolic potential model (Table 1). The relative weights of the two models were determined in a way to minimize the leave-one-out cross validation error (Table 1 and Supplementary Material section 4 [87]). We note that the weights determined this way, 0.75 for the cluster expansion and 0.25 for the parabolic potential model, are similar to what we would have obtained using inverse variance weighting [92] (0.72 for the cluster expansion and 0.28 for the parabolic potential model), using the leave-one-out cross validation errors to estimate the variance. The newly constructed weighted average of the cluster expansion and parabolic potential models further lowers the LOOCV error by 10% (13 meV) relative to cluster expansion itself (Figure S4, Supplementary Material [87]). This can be rationalized by the fact that these two models have error distributions that are largely uncorrelated due to their very different formalisms (Figure S5).

The improved performance of the cluster expansion comes with additional computational cost required to generate the training data. To assess the trade-off between cost and performance, we have evaluated the accuracy of the different methods as a function of training set size by randomly partitioning our data set into training sets and test sets, with models trained on the training sets and evaluated using the corresponding test sets. We started by randomly selecting 20% of the total data set as test set, training each model on all the remaining data, and then testing each model on the test set. This procedure was repeated 10 times to calculate the average root-mean-square test error. We then repeated the procedure using randomly-selected training sets that were 80%, 60%, 40%, and 20% as large as the first training set. The averaged test set root-mean-square errors for the 10 different runs at each size are shown in Figure 4, with the standard deviations colored as shaded regions.

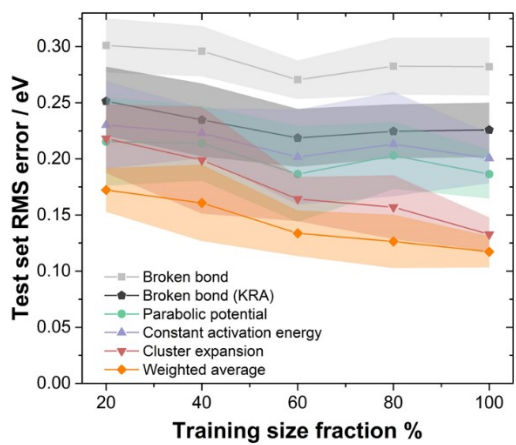

**Figure 4.** Test set root mean square (RMS) error as a function of training set size, expressed as a percentage of the remaining data set (excluding the test set), which are 234 structures. The test set contains 60 structures. The standard deviations of the test set errors are colored as shaded regions.

For the cluster expansion, the test set errors decrease significantly with increasing training set size, whereas for the broken bond models, constant activation energy model, and parabolic potential model the test set errors remain relatively constant (Figure 4). This can be understood by the more flexible form of the cluster expansion compared to the simpler models (Table 1). These results indicate that when little training data is available, the parabolic potential, constant activation energy, and KRA broken bond models can predict activation energies with accuracy comparable to the cluster expansion, but if higher accuracy is desired the cluster expansion can be improved through the generation of additional training data. The weighted average between the cluster expansion and parabolic potential model has the lowest error at every training set size and is an appealing option especially for relatively small training sets. For the system considered here, the weighted average model achieves the same accuracy of the cluster expansion with only 60% of the training data (Figure 4).

To investigate how the different models may affect the structural evolution of a Pt$_3$Ni nanoparticle and compare model execution speeds, we have performed KMC simulations on a truncated octahedral Pt$_3$Ni nanoparticle with randomly initialized atomic order (Pt$_{3411}$Ni$_{1137}$, with a diameter of approximately 6.2 nm) at 1000 K. The elevated temperature was chosen to accelerate dynamics and reduce the computational cost of the KMC simulations. In each KMC simulation, atoms were only allowed to hop into neighboring vacant sites and the activation energy was computed using different models. We also considered Ni dissolution, as significant Ni loss is observed experimentally [1,90,93]. We only allowed Ni dissolution from surface sites with a coordination number less than 10, and as long as all of the neighboring atoms were left with at least 3 nearest-neighbor atoms after Ni dissolution. For simplicity, for all models we assigned a zero activation energy for Ni dissolution, which generally occurs readily from the surface [1]. More details are provided in Supplementary Material section 2.2 [87]. The snapshots of the Pt-Ni nanoparticles after the KMC simulations are shown in Figure 5a-e.

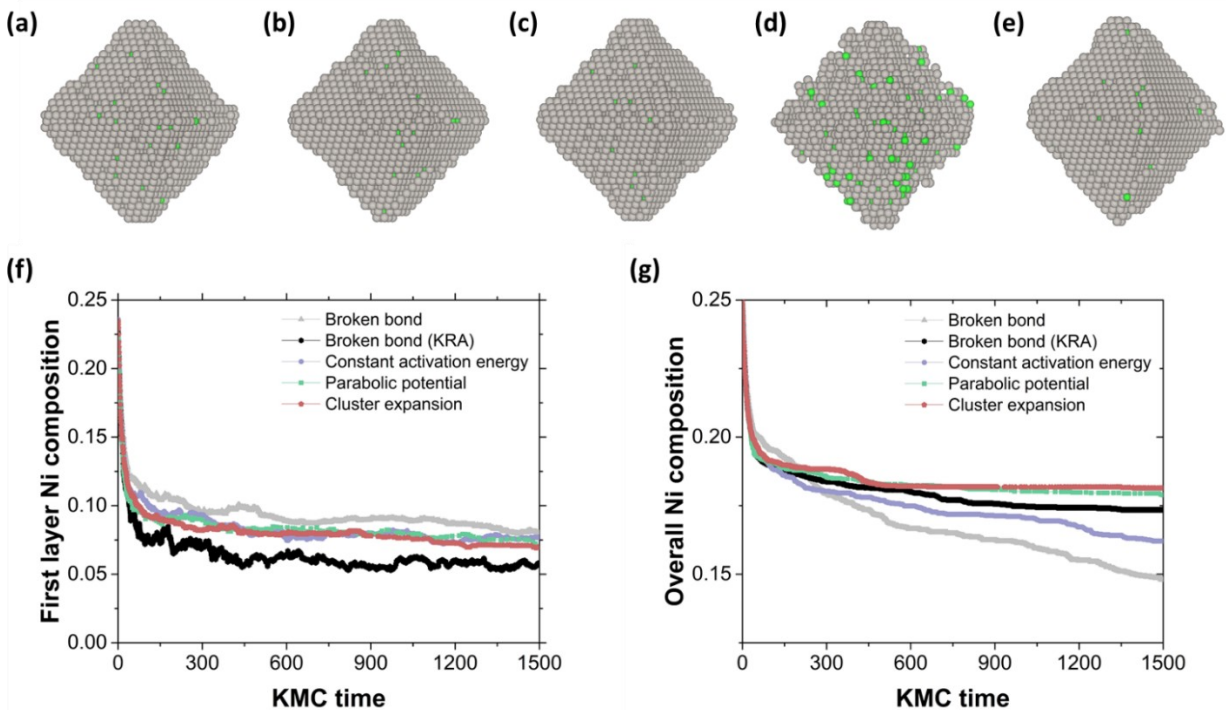

**Figure 5.** (a-e) Snapshots of Pt-Ni nanoparticles after the KMC simulations at 1000 K using the a) cluster expansion, b) parabolic potential model, c) constant activation energy model, d) simple broken bond model, and e) KRA broken bond model. Grey spheres represent Pt and green spheres represent Ni. (f, g) The first layer Ni composition and overall Ni composition from the KMC simulations as a function of KMC time.

The evolution of Ni composition in the first (outermost) layer is largely similar for the cluster expansion, parabolic potential model, and constant activation energy model (Figure 5f), where a Pt-rich shell is formed, consistent with our previous KMC work and experimental work [1,90]. The simple broken bond model produced a higher Ni composition in the first layer than that of the other models, but with the lowest overall Ni composition (Figure 5g), indicating that the Ni atoms are exchanged more rapidly from inner layers to the surface of the nanoparticle. The KRA broken bond model, however, shows completely different kinetics from the simple broken bond model, with a faster and more complete formation of a Pt shell and Pt(111) surface. This is likely due to the fact that the KRA broken bond model satisfies detailed balance, resulting in evolution towards a lower energy state.

The overall Ni dissolution rate (Figure 5g) largely follows the trend of the prediction errors of the five models. The simple broken bond model has the largest error and the cluster expansion has the smallest, with the parabolic potential and the constant activation energy model in between. The large prediction error of the simple broken bond model is responsible for promoting more hops with unfavorable change in energy, thus creating more defects and a more porous structure (Figure 5d). The less accurate models, such as the broken bond models and constant activation energy model, have less spread in the predicted activation energies and thus might artificially accelerate rare events that could lead Ni to segregate to the surface and dissolve, resulting in lower overall Ni compositions (Figure 5g).

Although the cluster expansion is much more accurate for predicting activation energies, it does not significantly affect the shape evolution of the $Pt_3Ni$ nanoparticle compared to the parabolic potential model or the constant activation energy model (Figure 5a-c, Figure 5f), except for the amount of predicted Ni loss. We analyze the nearest-neighbor environments of the Pt-Ni nanoparticles from the five models at the same overall Ni composition (18% Ni, Supplementary Material section 2.2 [87]). The values of $N_{Pt-Pt}$, $N_{Pt-Ni}$, and $N_{Pt-Vac}$, measurements of the average numbers of Pt-Pt bonds, Pt-Ni bonds, and Pt-vacancy bonds around a Pt atom, are almost identical for the cluster expansion, parabolic potential, and constant activation energy model, with the maximum difference being about 0.19 (Table S2). These values are also consistent with our previous KMC work using the constant activation energy model [1,90]. The KRA broken bond model has the largest value of $N_{Pt-Pt} + N_{Pt-Ni}$, consistent with the less loss of octahedral shape (Figure 5e). It is worth noting that although the parabolic potential model has a prediction error similar to that of the constant activation energy model, its Ni composition profile is much closer to that of the cluster expansion in the KMC simulations. This could be attributed to the more physically meaningful formalism of the parabolic potential model.

The simple broken bond model is the least accurate of the models, but it is much faster than the other models for calculating activation energies (by about 3 orders of magnitude) since it only evaluates number of nearest-neighbor bonds, which is equivalent to a cluster expansion with only nearest-neighbor pair ECI (Table 2). All other models calculate transition state energies roughly equally quickly, as they all use the complete cluster expansion, with 1097 ECI, to evaluate the energy difference between two different states (either the initial state and end state or the initial state and transition state).

**Table 2.** Average execution time for calculating the activation energy expressed relative to the time for the cluster expansion model.

|  | Relative time calculating activation energy |
|---|---|
| Broken bond | $1.673 \times 10^{-3}$ |
| Broken bond (KRA) | 1.210 |
| Parabolic potential | 1.175 |
| Constant activation energy | 1.160 |
| Cluster expansion | 1 |

The overall speed of the KMC simulation is not solely determined by the time required to calculate the activation energy. In the KMC routine the total time consists of the time i) choosing a random event, ii) checking if the event is valid according to its neighboring environment, iii) calculating the activation energy and transition probability if the event is valid, iv) updating the nanoparticle configuration if the event is accepted, and v) recording simulation data. Once calculating the activation energy becomes sufficiently fast, the remaining steps may become the bottleneck. In our implementation, due largely to the constraints on the allowed transitions on the surface, steps i) and ii) take up about as much time as calculating the activation energy using the cluster expansion, so methods that calculate the activation energy significantly more quickly than the cluster expansion have relatively little effect on the overall simulation time.

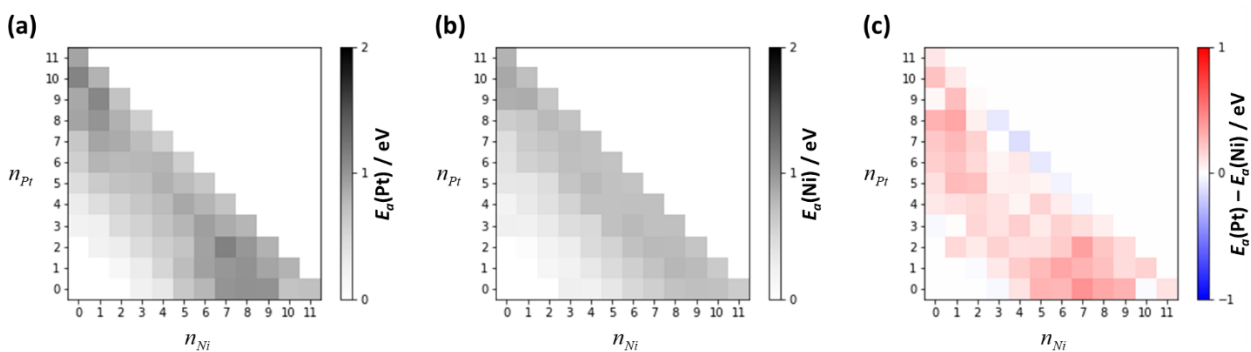

**Figure 6.** Cluster-expansion-predicted activation energies ($E_a$), averaged over 20,623 hops in randomly-generated nanoparticles, as a function of number of nearest-neighbor atoms $n_{Ni}$ and $n_{Pt}$ around the diffusing species. a) The activation energies for Pt. b) The activation energies for Ni. c) The differences between the activation energies in a) and b).

The transition state cluster expansion provides some insights into the activation energies for Pt and Ni diffusion as a function of the local environment. We randomly generated 100 nanoparticles with between 114 and 387 atoms and compositions ranging from 3% Ni to 97% Ni. The structures of these particles were determined in the same way as for the training structures. For each atom in each particle, we randomly generated a hop to a nearest-neighbor site (after making that site vacant if necessary), for a total of 20,623 random hops. The coordination number of the diffusing species in the initial site ranges from 3 to 11 in this data set. The average activation energies for Pt and Ni diffusion as a function of the nearest-neighbor environment are shown in Figure 6. The largest activation energies are not observed at the largest coordination number (11) as might be expected, but rather at coordination numbers of 10 or 9. This may be a reason for the relatively poor performance of the simple broken bond model, which cannot capture this non-linear trend. The largest activation energies occur near extreme Pt/Ni (or Ni/Pt) ratios, i.e. far away from 1:1, for both Pt diffusion and Ni diffusion. This may be due to the fact that the local environment has more structural degrees of freedom when it contains a mix of Pt and Ni atoms, providing more options for finding a low-energy path. Overall, we predict that the activation energy for Pt diffusion from undercoordinated sites is larger than that for Ni, but the activation energies are closer in the bulk (Figure 6c). The relatively slow Pt diffusion near the surface may play a role in protecting the underlying Ni atoms from dissolution [1,90,94,95].

## IV. CONCLUSIONS

We have presented and evaluated a cluster expansion model for predicting activation energies for vacancy-mediated diffusion in alloys by explicitly including sites representing saddle points on the potential energy surface. This approach allows for the calculation of activation energies using a single global cluster expansion and can be systematically improved with additional training data to the point at which it is roughly twice as accurate as commonly-used, simpler models, with comparable overall execution speed. Of the simpler models, the broken bond model is fastest at calculating activation energy but produces anomalous results due to its lack of detailed balance. A version of the broken bond model that corrects for this deficiency yields significantly improved results but is slower as it requires the evaluation of the energies of the initial and final states. Similar accuracy and speed are achieved by the constant activation energy model and the parabolic potential model. One benefit of the constant activation energy model is that it does not need to be trained on DFT-calculated activation energies to predict dynamics, as it calculates relative rates only using the difference in energy between the initial and final states. However, of the simpler models, the kinetic evolution of a Pt-Ni nanoparticle is most accurately represented by the parabolic potential model. This may be due to the better physical motivation behind the parabolic potential model, but the extent to which this result depends on the particular system being studied is not clear. A weighted average of the cluster expansion model and parabolic potential model is most accurate at all training set sizes, suggesting this may be an effective approach in general.

## ACKNOWLEDGEMENTS

This work was supported by the Office of Naval Research (Grant No. ONR MURI N00014-15-1-2681). C.L. and T.M. acknowledge the computational resources provided by the Brookhaven National Laboratory under grant 33818, the Maryland Advanced Research Computing Center (MARCC), and the Homewood High Performance Cluster (HHPC). The authors also thank Alberto Hernandez for helpful discussions. Atomic-scale structural images were generated using VESTA [96].

## APPENDIX

In this appendix we will prove that if configuration space is constrained to prevent the simultaneous occupation of two sites, then all cluster functions that depend on the occupancies of both of those sites can be removed from the expansion and the remaining functions form a complete basis.

We start by introducing some background and notation. Let the number of possible states for the $j^{th}$ site of a cluster expansion be given by $N_j$. For example, in the Pt-Ni-Vacancy cluster expansion, $N_j = 3$ for all sites, as each site can be occupied by a vacancy, Pt atom, or Ni atom. For a given configuration, let the occupation of the $j^{th}$ site be indexed by the site variable $s_j$, which can take on values from 1 to $N_j$. For example, we may have $s_j = 1$ if the $j^{th}$ site is vacant, $s_j = 2$ if the $j^{th}$ site is occupied by Pt, and $s_j = 3$ it the $j^{th}$ site is occupied by Ni. The set of site variables for all sites for a given atomic configuration is given by the vector $\mathbf{s}$. Here we will always define the site variables so that $s_j = 1$ whenever the $j^{th}$ site is vacant.

To construct the cluster expansion basis, we select a set of $N_j$ independent basis functions, $\Theta_i(s_j)$, for each site, where $i \in \{1, 2, ..., N_j\}$. Let one of these basis functions be the constant value, 1. We will always assign this value to the first basis function, so for every site we have $\Theta_1(s_j) = 1$. The tensor product of all of these single-site bases define a complete basis of "cluster functions" that can be used to represent any function of $\mathbf{s}$. Every one of the cluster functions in this basis is a product of a unique combination of single-site basis functions, with the product including exactly one basis function for each site:

$$\Phi_{\mathbf{b}}(\mathbf{s}) = \prod_j \Theta_{b_j}(s_j) , \tag{1}$$

where the vector $\mathbf{b}$ represents the particular combination of single-site basis functions that define this cluster function, and $b_j$ is the index of the single-site basis function for the $j^{th}$ site. As $\Theta_1(s_j) = 1$, each cluster function is a function of only the site variables for which $b_j \neq 1$ [32]. The corresponding sites are the "cluster" represented by the cluster function, and we say the cluster function "includes" these sites.

For each site, the allowed values for $b_j$ are identical to the allowed values for $s_j$, as they are both integers between 1 and $N_j$. As the number of possible configurations is given by the number of possible unique vectors $\mathbf{s}$, and the number of basis functions is given by the number of possible unique vectors $\mathbf{b}$, we can see that the number of basis functions equals the number of total possible configurations. I.e. for any configuration $\mathbf{s}$, we can identify a unique basis function for which $b_j = s_j$ for all $j$, and vice versa.

Now we consider the case where there are two sites that cannot be occupied at the same time. We will let $j = 1$ for the first site and $j = 2$ for the second. The constraint that both sites cannot be occupied eliminates all configurations for which $s_1 \neq 1$ and $s_2 \neq 1$. As a result, there are now more cluster functions than possible configurations.

We can eliminate redundant cluster functions by eliminating all cluster functions that are functions of $s_1$ and $s_2$, i.e. cluster functions for which $b_1 \neq 1$ and $b_2 \neq 1$. The number of remaining vectors $\mathbf{s}$ is equal to the number of remaining vectors $\mathbf{b}$, as for any configuration $\mathbf{s}$, we can still identify a unique basis function for which $b_j = s_j$ for all $j$, and vice versa. Thus, the number of remaining cluster functions equals the number of allowed configurations after imposing the constraint.

To show that the remaining cluster functions form a complete basis, we need only show that they are linearly independent. We do this by demonstrating that each of the cluster functions we have removed can be expressed as a linear combination of cluster functions that were not removed. Specifically, let a cluster function that includes sites 1 and 2 be labeled by

$$\mathbf{b}_{++} = \{b_1, b_2, b_3, b_4, ...\}, \tag{2}$$

where $b_1 \neq 1$ and $b_2 \neq 1$. We similarly define the labels

$$\begin{aligned}\mathbf{b}_{-+} &= \{1, b_2, b_3, b_4, ...\}, \\ \mathbf{b}_{+-} &= \{b_1, 1, b_3, b_4, ...\}, \\ \mathbf{b}_{--} &= \{1, 1, b_3, b_4, ...\}\end{aligned} \tag{3}$$

for cluster functions in which site 1 is removed from $\Phi_{\mathbf{b}_{++}}(\mathbf{s})$, site 2 is removed from $\Phi_{\mathbf{b}_{++}}(\mathbf{s})$, and both sites 1 and 2 are removed from $\Phi_{\mathbf{b}_{++}}(\mathbf{s})$, respectively. As none of these three cluster functions include both sites 1 and 2, none will be removed from the expansion. For the allowed values of $s_1$ and $s_2$, $\Phi_{\mathbf{b}_{++}}(\mathbf{s})$ can be written as a linear function of $\Phi_{\mathbf{b}_{-+}}(\mathbf{s})$, $\Phi_{\mathbf{b}_{+-}}(\mathbf{s})$, and $\Phi_{\mathbf{b}_{--}}(\mathbf{s})$:

$$\begin{aligned}\Phi_{\mathbf{b}_{++}}(\mathbf{s}) &= \Theta_{b_1}(s_1)\Phi_{\mathbf{b}_{-+}}(\mathbf{s}) \\ &+ \Theta_{b_2}(s_2)\Phi_{\mathbf{b}_{+-}}(\mathbf{s}) \\ &- \Theta_{b_1}(s_1)\Theta_{b_2}(s_2)\Phi_{\mathbf{b}_{--}}(\mathbf{s})\end{aligned} \tag{4}$$

If there are other pairs of sites that cannot be simultaneously occupied, then the procedure can be repeated.

Supplementary Material

# Predicting activation energies for vacancy-mediated diffusion in alloys using a transition-state cluster expansion

Chenyang Li,[1] Thomas Nilson,[1] Liang Cao,[1] and Tim Mueller[1, *]

[1]Department of Materials Science and Engineering, Johns Hopkins University, Baltimore, MD 21218, USA

*Email: tmueller@jhu.edu

## 1. Cluster expansion

A ternary cluster expansion was built based on an fcc lattice, where each site can be occupied by platinum (Pt), nickel (Ni), or a vacancy. We define a set of sublattice sites located at halfway between two neighboring fcc lattice sites to represent all possible transition states for vacancy-mediate atomic diffusion on the fcc lattice (we will refer to these as "transition state sites"). In a cluster expansion the occupancy of the site $i$ can be represented by a site variable $s_i$. For ternary cluster expansion in this work, site variable values of 1, 2, and 3 were assigned to vacancy, Pt, and Ni, respectively, and we used a discrete cosine basis to generate the cluster functions.

For fitting the cluster expansion, a training set with a total number of 299 relaxed structures was generated using DFT. The training set contained the empty structure, bulk Pt, bulk Ni, bulk PtNi, bulk $Pt_3Ni$, and sets of nanoparticles (ranging from 90 to 226 atoms) representing diffusive hops with the initial, transition, and final state. Each of the bulk structures was included twice to ensure accuracy in the bulk limit. The structures of the nanoparticles were randomly generated by running a Monte Carlo simulation [1] at 2000 K for 184,320 steps, where the ECI for nearest-neighbor pairs was set to -1 eV and all other ECI were set to 0. The last 84 structures in the training set were randomly generated surface hops, which are important to the structural evolution of the nanoparticle. In this paper we considered one single hopping atom per supercell. To eliminate degeneracy in the cluster expansion, we removed all clusters which contain both a transition state site and at least one of the neighboring fcc sites, as well as cluster that contain more than one transition state site. The cluster expansion contained 1097 symmetrically distinct cluster functions and the ECIs for these cluster functions were fit to the DFT-calculated training set using the Bayesian approach with a multivariate Gaussian prior distribution [2]. The inverse of the covariance matrix for the Gaussian prior was diagonal, with elements given by:

$$\lambda_{\alpha\alpha} = \begin{cases} 0, & n_\alpha = 0 \\ e^{-\lambda_1}, & n_\alpha = 1, \text{ include transition state} \\ e^{-\lambda_2}, & n_\alpha = 1, \text{ not include transition state} \\ e^{-\lambda_1} e^{\lambda_3 r_\alpha} e^{\lambda_4 n_\alpha}, & n_\alpha > 1, \text{ include transition state} \\ e^{-\lambda_2} e^{\lambda_3 r_\alpha} e^{\lambda_4 n_\alpha}, & n_\alpha > 1, \text{ not include transition state} \end{cases}$$

where $n_\alpha$ is the number of sites in cluster function $\alpha$, and $r_\alpha$ is the maximum distance between sites. The $\lambda_1$, $\lambda_2$, $\lambda_3$, and $\lambda_4$ were determined using a conjugate-gradient algorithm to minimize the root mean square leave-one-out cross validation (LOOCV) score, which is an estimate of the prediction error [3]. Final values of the parameters $\lambda_1$, $\lambda_2$, $\lambda_3$, and $\lambda_4$ for the Pt-Ni-vacancy cluster expansion are 20.843, 19.990, 1.787, 1.849, respectively. The cluster expansion has a LOOCV error of 0.127 eV for the activation energies and 1.203 meV/atom for the formation energies relative to DFT calculations.

## 2. Monte Carlo simulations

*2.1. Metropolis Monte Carlo.* Simulated annealing was used to determine the equilibrium shape and layer compositions of the $Pt_3Ni$ nanoparticle ($Pt_{2547}Ni_{849}$). A Metropolis Monte Carlo simulation in a canonical ensemble [1] was run from a high temperature (2000 K), and then decreased in steps by a factor of $4^{0.10}$ until 333 K. At each temperature, the number of Monte Carlo iteration was 20 times the number of sites in the supercell. The supercell was 62.4 Å × 62.4 Å × 62.4 Å. The thermodynamically averaged layer compositions were recorded during the Monte Carlo sampling at 333 K.

*2.2. Kinetic Monte Carlo.* Kinetic Monte Carlo (KMC) [4,5] simulations were used to study the structural evolution of the $Pt_3Ni$ nanoparticle, where atoms were only allowed to hop into neighboring vacant sites. We only allowed Ni dissolution from a surface site with a coordination number less than 10 and all of its neighboring atoms having more than 2 nearest-neighbor atoms after Ni dissolution, and we assigned a zero activation energy for Ni dissolution due to the fast dissolution rate observed from experiments [6]. The activation energies for atomic diffusion were computed using different methods outlined in Table 1. We used the standard rejection KMC algorithm:

a) Start at time $t = 0$ with an initial state $i$ and compute the number of all possible transition events $N_i$ (diffusion and dissolution).

b) Randomly choose an event by uniformly sampling the $N_i$ events and accept the event with a probability $k_{ij}/k_0$, where $k_{ij} = v\exp(-E_a/k_B T)$ is the transition rate from state $i$ to $j$, $E_a$ is the activation energy, $v$ is the prefactor, and $k_0$ is an upper bound for the transition rate. We set $k_0 = v$ which corresponds to a zero activation energy, and we used the same prefactor for diffusion and dissolution.

c) Update KMC time by $t := t + \Delta t$, where $\Delta t = (N_i k_0)^{-1} \ln(1/u)$, and $u$ is a uniform random variable from $(0,1]$. We used an upper bound of $k_0 = 1$ to determine "KMC time".

d) Go to step b).

We performed KMC simulations on a truncated octahedral Pt₃Ni nanoparticle with random initial atomic order (Pt$_{3411}$Ni$_{1137}$, with a diameter of approximately 6.2 nm). The supercell was 74.1 Å × 74.1 Å × 74.1 Å. We chose an elevated temperature of 1000 K to accelerate the dynamics and reduce the computational cost of the KMC simulations. We stopped the KMC simulations when the KMC time reached 1500. We calculated the nearest-neighbor environments for the Pt-Ni nanoparticles, using the equations (5) below. The values for $N_{Pt-Pt}$, $N_{Pt-Ni}$, and $N_{Pt-Vac}$ are measurements of the average number of Pt-Pt bonds, Pt-Ni bonds, and Pt-vacancy bonds around a Pt atom ($N_{Ni-Pt}$, $N_{Ni-Ni}$, and $N_{Ni-Vac}$ are for the Ni atom similarly). The results averaged over five runs are provided in Table S2.

$$N_{Pt-Pt} = \frac{\#bond_{Pt-Pt}}{\#atom_{Pt}}, N_{Pt-Ni} = \frac{\#bond_{Pt-Ni}}{\#atom_{Pt}}, N_{Pt-Vac} = \frac{\#bond_{Pt-Vac}}{\#atom_{Pt}} = 12 - N_{Pt-Pt} - N_{Pt-Ni}$$
$$N_{Ni-Pt} = \frac{\#bond_{Ni-Pt}}{\#atom_{Ni}}, N_{Ni-Ni} = \frac{\#bond_{Ni-Ni}}{\#atom_{Ni}}, N_{Ni-Vac} = \frac{\#bond_{Ni-Vac}}{\#atom_{Ni}} = 12 - N_{Ni-Pt} - N_{Ni-Ni}$$
(5)

*2.3. Constant activation energy model.* We give a brief mathematical proof of how the value of the constant activation energy only affects the prefactor of the KMC simulation. The functional form is:

$$E_a = \begin{cases} c, & \Delta E < 0 \\ c + \Delta E, & \Delta E \geq 0 \end{cases}.$$

Therefore, the transition rate can be calculated:

if $\Delta E < 0$, $k_{ij} = v\exp(-E_a/k_BT) = v\exp(-c/k_BT) = vk^*\exp(-0/k_BT) = vk^*$,
where $k^* = \exp(-c/k_BT)$.
if $\Delta E \geq 0$, $k_{ij} = v\exp(-E_a/k_BT) = v\exp(-(c+\Delta E)/k_BT) = vk^*\exp(-\Delta E/k_BT)$,

where $k^* = \exp(-c/k_BT)$. The prefactor $vk^*$ only affects the time scale of the simulation and can be set to an arbitrary value without affecting the dynamics, which depend only on $\Delta E$. Thus, it is possible to run KMC simulations using this model without having to train a transition-state cluster expansion.

## 3. Illustration of the parabolic potential model

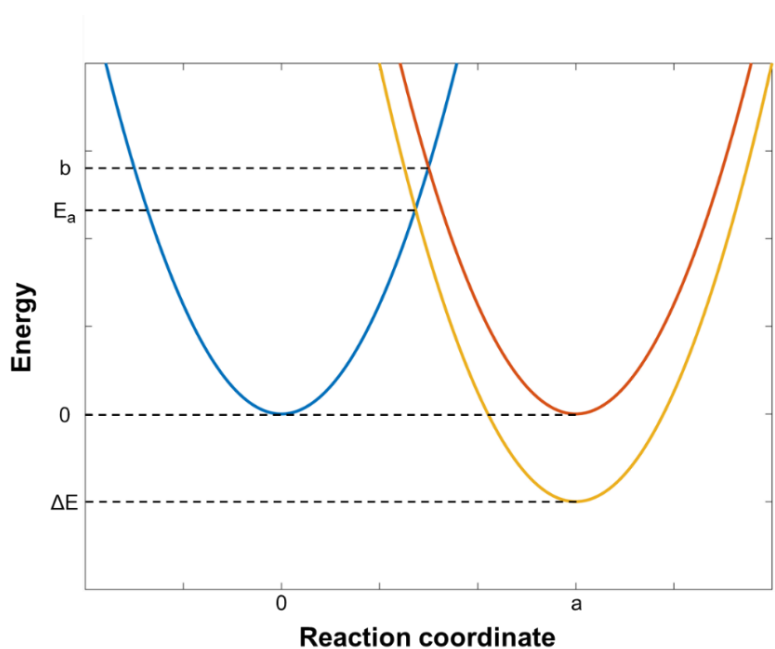

**Figure S1.** Illustration of the parabolic potential model. The blue curve is the initial state, the red curve is the final state with the same energy, and the yellow curve is the final state with a lower energy $\Delta E$. The intersection of two parabolas indicates the energy of the transition state.

The following derivation is adapted from Xiao and Henkelman [7]. Assume the parabola takes the form of $E = kx^2$, where $x$ is the reaction coordinate. We have:

$$b = k\left(\frac{a}{2}\right)^2 \tag{6}$$

For the end state with a change in energy $\Delta E$, we solve for the new activation energy $E_a$ and the reaction coordinate $x = r$:

$$E_a = kr^2 = k(r-a)^2 + \Delta E. \tag{7}$$

We obtain the activation energy to be:

$$E_a = \frac{(\Delta E + 4b)^2}{16b}. \tag{8}$$

## 4. Weighted average model

The relative weights for the base models were determined to minimize the root mean square error (equation 9) using the cross-validation value:

$$\sqrt{\frac{1}{N}\sum_{i=1}^{N}\left(y_i - \sum_{j=1}^{n} w_j \hat{y}_{ij}\right)^2},$$

$$s.t. \sum_{j=1}^{n} w_j = 1,$$

(9)

where $N$ is the number of entries, $n$ is the number of base models, $w_j$ is the weight for the $j^{th}$ model, $\hat{y}_{ij}$ is the cross-validation value of entry $i$ from the $j^{th}$ model, and $y_i$ is the true (DFT) value of entry $i$.

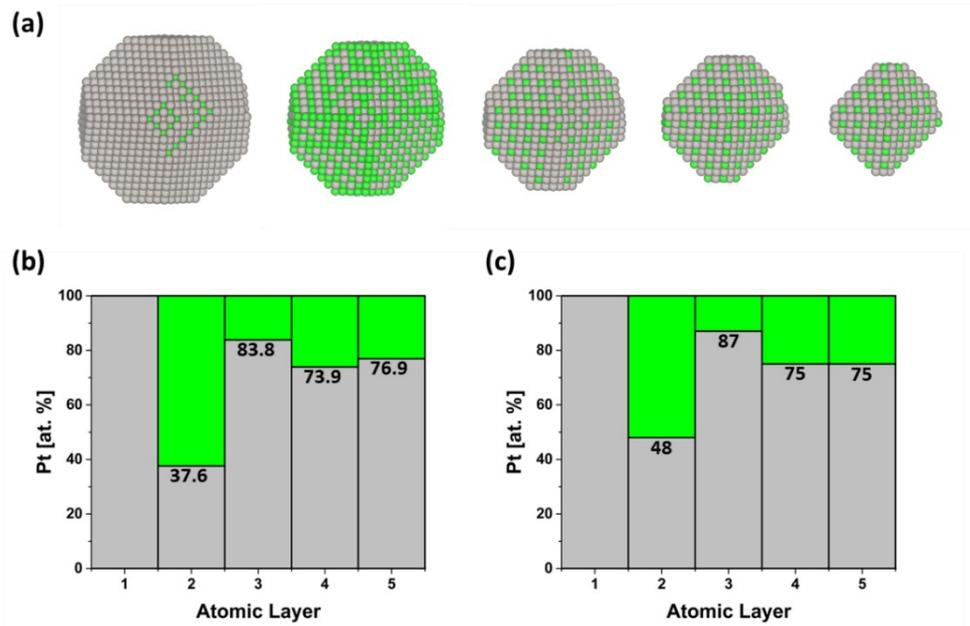

**Figure S2.** (a) Metropolis Monte Carlo snapshot of a $Pt_{2547}Ni_{849}$ nanoparticle at 333 K, showing the first, second, third, fourth, and fifth layer, from left to right. Gray and green spheres represent Pt and Ni, respectively. (b) Thermodynamically averaged surface composition profile for a $Pt_{2547}Ni_{849}$ nanoparticle at 333 K. (c) Surface composition profile for a $Pt_3Ni(111)$ surface from experiments at 333 K [8].

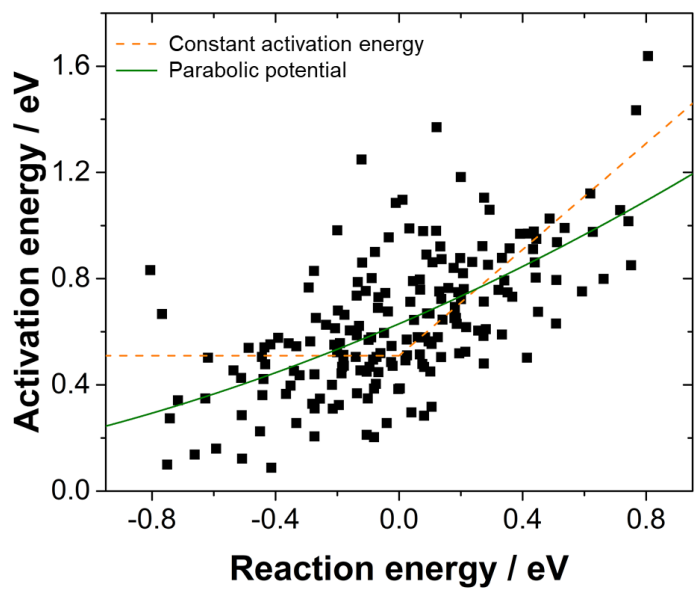

**Figure S3.** DFT activation energies plotted against the reaction energies Δ*E* (energy differences between final state and initial state) for the entire data set. The orange and green lines are the best fits using the constant activation energy model and the parabolic potential model, respectively.

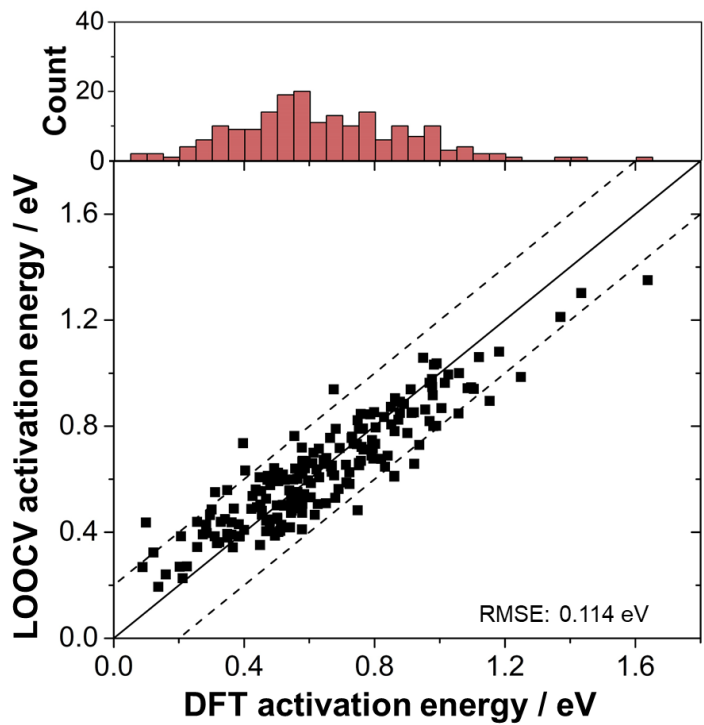

**Figure S4.** Leave-one-out cross validation (LOOCV) of the activation energies from the weighted average model against the known DFT activation energies. The dashed lines indicate ± 0.2 eV deviation from perfect agreement. The upper panel shows the distribution of activation energies in the training set.

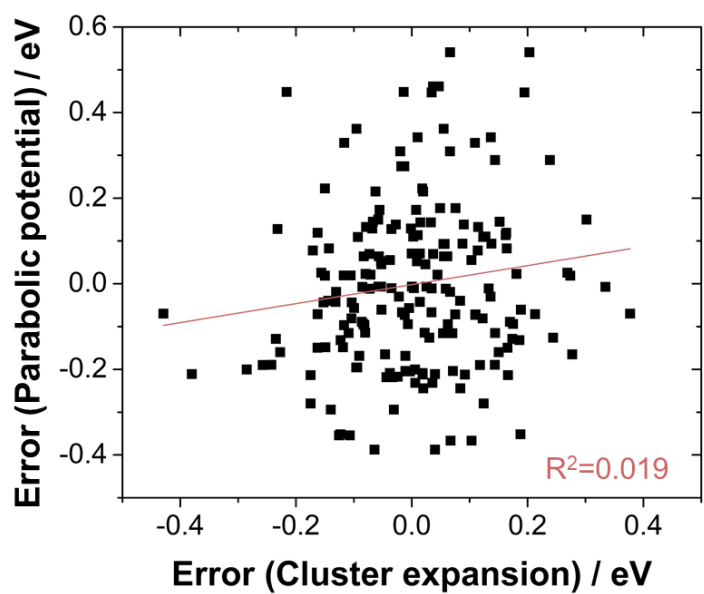

**Figure S5.** Scatter plot of leave-one-out cross validation errors for the cluster expansion and the parabolic potential model, which shows the lack of correlation.

**Table S1.** Various models for predicting activation energies parameterized from ab initio data. The reported errors are root-mean-square errors (RMSE) unless otherwise noted.

| System | Approach | Fitting error / meV | Validation error / meV | Ref. |
| --- | --- | --- | --- | --- |
| Vacancy-mediated diffusion in Pt-Ni nanoparticles | CE | 83 | 127 | This work |
| Lithium diffusion in $Li_xCoO_2$ | CE | 40 | - | [9] |
| FCC solute diffusion (5 hosts) | ANN/GKRR | 92~105 | 148~154 | [10] |
| Impurity diffusion (15 hosts) | GPR/GKRR | - | 116~155 | [11] |
| Vacancy diffusion in Fe-Cu alloys | GPR | - | 67[a] | [12] |
| Lithium diffusion in $Li_3PO_4$ | ANN | 48~73[a] | - | [13] |
| Self-diffusion on Cu surfaces | ANN | ~161 | - | [14] |
| Transition metal solute diffusion | SVM | 92 | 142 | [15] |
| Interstitial diffusion of nitrogen, oxygen, boron, and carbon in metals | GB | 135 | 311 | [16] |
| Cu bulk vacancy diffusion | SR | - | 37~106 | [17] |

CE: cluster expansion; ANN: artificial neural network; GPR: Gaussian process regression; GKRR: Gaussian kernel ridge regression; SVM: support vector machine; GB: gradient boosting; SR: symbolic regression.

[a] mean absolute error.

**Table S2.** The nearest-neighbor coordination environments of the Pt-Ni nanoparticles averaged over five simulations at an overall Ni composition of 18% ($Pt_{3411}Ni_{748}$). $N_{A-B}$ represents the average number of nearest-neighbor bonds around atom "A".

|  | $N_{Pt-Pt}$ | $N_{Pt-Ni}$ | $N_{Pt-Vac}$ | $N_{Ni-Pt}$ | $N_{Ni-Ni}$ | $N_{Ni-Vac}$ |
|---|---|---|---|---|---|---|
| Broken bond | 7.98 | 1.91 | 2.11 | 8.73 | 2.56 | 0.71 |
| Broken bond (KRA) | 8.34 | 2.00 | 1.66 | 9.13 | 2.62 | 0.25 |
| Constant activation energy | 8.17 | 1.96 | 1.87 | 8.93 | 2.72 | 0.35 |
| Parabolic potential | 8.15 | 1.96 | 1.89 | 8.95 | 2.70 | 0.35 |
| Cluster expansion | 8.31 | 1.99 | 1.70 | 9.09 | 2.63 | 0.28 |